\documentclass[twocolumn]{elsart}

\usepackage{graphicx}

\usepackage{natbib}

\def\simlt{\hbox{ \rlap{\raise 0.425ex\hbox{$<$}}\lower 0.65ex\hbox{$\sim$} }}
\def\simgt{\hbox{ \rlap{\raise 0.425ex\hbox{$>$}}\lower 0.65ex\hbox{$\sim$} }}

\begin{document}
\begin{frontmatter}

\noindent

\title{Dense Stellar Systems as Laboratories for Fundamental Physics}
\author{Piet Hut}\ead{piet@ias.edu}

\address{Institute for Advanced Study, Princeton, NJ 08540, USA}

\begin{abstract}
Galactic nuclei and globular clusters act as laboratories in which
nature experiments with normal stars, neutron stars and black holes,
through collisions and through the formation of bound states, in the
form of binaries.  The main difference with the usual Earth-based
laboratories is that we cannot control the experiments.  Instead, we
have no choice but to create virtual laboratories on Earth, in order
to simulate all the relevant physics in large-scale computational
experiments.  This implies a realistic treatment of stellar dynamics,
stellar evolution, and stellar hydrodynamics.

Each of these three fields has its own legacy codes, workhorses that
are routinely used to simulate star clusters, stars, and stellar
collisions, respectively.  I outline the main steps that need to be
taken in order to embed and where needed transform these legacy codes
in order to produce a far more modular and robust environment for
modeling dense stellar systems.

The time is right to do so: within a few years computers will reach
the required speed, in the Petaflops range, to follow a star cluster
with a million stars for ten billion years, while resolving the
internal binary and multiple star motions.  By that time simulation
software will be the main bottleneck in our ability to analyze dense
stellar systems.  Only through full-scale simulations will we be able
to critically test our understanding of the `microphysics' of stellar
collisions and their aftermath, in a direct comparison with observations.

\end{abstract}

\end{frontmatter}

\section{Gravitational Laboratories}

In experimental high-energy physics, there are two ways to probe
elementary particles.  One way is by studying the properties of their
bound states, which may be more stable or more easily accessible than
unbound particles.  The other way is by smashing particles together,
and to study the remnants emerging from such collisions.  In this way,
we have learned an enormous amount in the last century about the
electroweak and strong interactions.  Gravity, however, has remained
almost totally elusive.

The problem is related to the weakness of the gravitational force,
which implies that we have to add quite a number of particles before
gravity can dominate.  A self-gravitating bound state of nuclear
matter, a neutron star, contains some $10^{57}$ particles, in a ball
with a diameter of order 10 km, and a mass like that of the sun.  And
the only known way to produce a purely gravitational object, a black
hole, of moderate size, is to let a star implode to form an object
of several solar masses with a horizon size of the order of ten
kilometers or so.  Alas, we have no laboratories in which to create
such objects.

Fortunately, nature has been kind enough to provide us with labs in
the sky.  And we don't have to look far away, either.  We are
accustomed to observe quasars and gamma ray bursts at distances
measured in Gigaparsecs, but our local gravitational laboratories
are a million times closer than that.  There are dozens of them
in our very own Milky Way galaxy, close enough to have a good peek to
see what is going on.

At a distance of several kiloparsecs, a number of globular clusters
have a high enough central density to let neutron stars interact and
collide with other stars, often forming exotic binaries as byproducts.
And at a distance of less than ten kiloparsecs, our galactic nucleus
contains the mother of all gravitational laboratories, where a central
massive black hole of a few million solar masses is surrounded by
swarms of neutron stars, black holes and all kind of stars, some of
them very unusual looking and all of them prone to collisions.

So here we are, with great gravitational laboratories, just around
the corner of where we live, cosmologically speaking, and we are
accumulating a treasure trove of high-precision observational data.
The main problem in analyzing the data is that we came in late.
Most of the experimental runs that we are currently watching were
started at least millions and sometimes billions of years ago.  In
order to interpret the data correctly, we have to reconstruct what has
happened during the time that the experiments have been underway.

For some purposes, we can make back-of-the-envelope estimates to
describe the essence of some of the main physical processes involved.
For more detailed investigations, however, we have no choice but to
conduct computer simulations in which we reconstruct the history of
the stellar system under consideration.  In many astrophysical
simulations, this task neatly breaks up into the task of simulating
the individual elements, such as stars and binaries on the one hand,
and the star system on the other.  In the case of dense stellar
systems, however, by definition such a clear separation is not
possible.

Dense stellar systems are characterized by an ecological network,
where everything influences everything.  In a globular cluster, for
example, dynamical interactions between passing stars can form new
binaries and modify the properties and even the membership of existing
binaries.  At the same time, internal changes in binaries, through
mass overflow or coalescence, feed back into the energy budget of the
star cluster as a whole.  In fact, most globular clusters probably
have more gravitational binding energy locked up in internal binary
degrees of freedom than in the bulk binding energy of the cluster as a
whole.  Even a partial access to those internal `microscopic' degrees
of freedom can greatly influence the `macroscopic' behavior of a star
cluster as a whole.

Progress in our understanding of dense stellar systems is therefore an
extreme form of two-step process.  We can gain a considerable insight
in the basic processes that are at work, through dimensional analysis
and back-of-the-envelope calculations, as was done very successfully
in the seventies and to some extent in the eighties.  Having
identified the main processes, we then suddenly faced a wall: in order
to make significant further progress, we had no choice but to model
the whole ecological network.  This has proved daunting: we are still
only in the initial phases of living up to this challenge.  The
current paper provides an outline of how we will plunge into this
problem fully during the next ten years.

\section{Complexity}

It sounds so simple: in order to make a large-scale simulation of a
system of interacting stars, why not just hook together existing
codes, each of which takes care of some of the physics?  After all, the
first stellar evolution codes were written in the fifties, and the
first stellar dynamics codes in the sixties.  By the seventies, both
fields were reaching a degree of maturity, and also the first
simulations of stellar collisions were carried out.  In the thirty
years since then, computer speed has increased a million-fold.  What
we could do on the level of stellar evolution for a few stars back in
the seventies, we should now be able to do easily for a million stars.
What is holding us back?

The answer can be given in one word: complexity.

The one single bottleneck in modern-day technology is software.  While
hardware is getting faster, we seem to be almost lost with respect to
the task of writing software to make good use of this speed.  The
bottom line is that we have not yet learned some of the basic
principles of good software writing.

One fundamental problem which we simply don't know how to handle yet
is pattern recognition.  Letting a computer recognize the identity of
an individual human face from a photograph is notoriously difficult,
even for a computer that executes far more elementary operations per
second than the human brain fires neurons per second.  Some progress
is being made, using more or less brute force, but clearly we haven't
yet found an efficient way of dealing with this challenge.

An example where brute force did work is chess, where computers can now
easily beat the strongest players.  But a problem such as go (a
traditional board game, {\it weichi} in Chinese, {\it (i)go} in
Japanese, {\it baduk} in Korean), has also attracted considerable
attention, but with very little success: even a beginning go player,
having practiced the game for only a few months, has a good chance to
beat the world's strongest go playing program.

Another fundamental problem is scalability.  Once we have written a
single computer code, or a whole software package, the challenge to
let these codes grow to cover a more complex situation is enormous.
Additionally the need for more CPU-cycles will require to distribute
the execution over a network, which is extremely difficult to
accomplish, if the system is not designed as a distributed system,
which is often not (or only partly) the case.  As a result, software
projects are almost always over time, over budget, contain lots of
bugs, are not sufficiently compatible with other programs or even with
themselves; the list of grievances goes on and on.

Here, too, it seems pretty clear that we are still lacking insight in
some fundamental principles, yet to be discovered.  And this is not
surprising.  A computer is fundamentally different from other tools
that we have built over the millennia.  An airplane or a spaceship,
while a modern invention, still resembles in some ways a boat, and
there is an almost continuous evolution in vehicles of transportation,
from horse-drawn carriages to trains and automobiles to planes and
spacecraft.  The switch from an abacus to a computer, however, is
qualitatively different.

An abacus is a pure object, manipulated by a human.  A computer, in
contrast, contains a program that effectively turns the machine into a
type of subject, an autonomous agent, manipulating itself.  Who knows,
it may well take another half century before we understand how to
operate this new type of tool on a deep enough level to perform
efficient pattern recognition and scaling; all we know is that we
haven't succeeded yet in the last half century.

In the specific case of computational astrophysics, we find a
situation that is even more problematic that it is for software
development in general.  Of course we do not really know yet how to
develop software, something that is true for all fields.  But what is
worse, it seems that in astrophysics by and large the community does
not even know that we do not know.  As a result, students in
astrophysics learn a lot about theory and observations, but preciously
little about simulations.

\section{Simulations as the Third Pillar of Science}

Astrophysics is not the only area in science where simulations have
been neglected in the standard curriculum.  The main problem seems to
be that computer simulations still tend to be seen as part of theory,
in a dichotomy between theory and experiment, or in the case of
astronomy, a dichotomy between theory and observation.

It is true that the earliest computer calculations were more or less
an extension of pen-and-paper calculations, albeit millions of times
faster already in the nineteen sixties.  But by now, computers are a
billion times a million times faster than humans, and this
quantitative growth has definitely made a qualitative difference.
While setting up a physics simulation still resembles work in
theoretical physics, analyzing the results of a simulation has much
more in common with experimental or observational physics.  And
writing the software for a sophisticated multi-scale multi-physics
simulation environment is not that different from designing an equally
sophisticated laboratory or telescope.

Clearly, then, a standard education program in (astro)physics should
contain training in all three prongs of modern physics: theory,
experiment, and simulation.  Nobel laureate Ken Wilson was one of the
first physicists to make a clarion call, a quarter century ago, by
describing computer simulations as the third paradigm of science,
but his message still has not come across clearly.  It is interesting
to ask why this is so; if we understand the reasons better, we might
be in a better position to do something about it.

I think the main problem is the magnitude of the cultural step
required to accept simulations as a true pillar of natural science.
Computational science is not a branch of science, but rather a pillar.
We would not call theory a `branch of science', nor would we call
laboratory work a `branch'.  And when we look back at history, we see
that accepting a new `pillar' has always taken a few generations.

The Greeks started theory, in full form with Euclid's axiomatic
approach, more than two thousand years ago.  However, the Greeks
lacked an equal appreciation for experiments, and it was only around
the time of Galileo, four hundred years ago, that experiment and
theory were joined in the conception of modern science.  I am not an
expert in the history of science, but I am sure that this addition of
a second pillar must have taken a few generations before everything
was shaken down comfortably.

The idea of keeping a note book for your lab experiments, and sharing
what you have found openly and in all necessary detail with others,
must have been very different from the approach of the Medieval guilds
whose members kept their trade secrets to themselves.  The scientific
program, in contrast, has been `open source' from the beginning.  And
this has been the great strength of science, that new ideas could be
immediately tested by anyone capable of doing so, a process that has
led to rapid and robust progress.

We are currently in the middle of a transition to adding simulations
as a third pillar to science, and I expect that it will take a few
generations before we have found a comfortable and generally accepted
way to make all three pillars fit together well.  Perhaps by 2050 we
will look back at the current confusion and wonder what took us so
long.  The answer is: it takes much longer to change the {\it way} in
which you do something than {\it what} you are doing.

\section{The Need for a Whole New Approach}

In astrophysics, and in physics in general, almost all computer
simulations have been carried out in the same mode as theoretical
investigations, as largely individual endeavors.  While there may be
collaborations in writing papers, by and large a single person is
responsible for writing the software for a simulation.  After several
years, someone else may take over such a code, and develop it further,
but at any given time, typically only one person is actively
developing the software.

If we compare this approach to that of experimentation or observation,
we see a striking difference.  Two hundred years ago it may well have
been possible to build your own telescope, but already a hundred years
ago that would have been out of the question.  Observation is team work:
no individual can possibly prepare the infrastructure for even a single
cutting-edge telescope at a single site.

It is inevitable that the same will become true for simulations.  In
the case of experimentation and observation, the complexity of the
hardware grew in such a way that a single person could no longer take
care of all of it.  And in the case of simulations, the complexity of
the software is rapidly reaching that same point of no return: within
the next ten years, no major field of computational astrophysics will
remain simple enough for a graduate student to write a complete
simulation code from scratch, something that is currently still
(barely) doable in some fields.

This does not mean that the inspiration for a new approach cannot come
from a single individual.  On the contrary.  Most successful
designs of novel telescopes or novel detectors have embodied the vision
of a leading expert, with enough experience and inspiration to convince
others to collaborate to work out the vision in all necessary detail,
from funding to construction to maintaining the whole operation.  But
even if the inspiration can be traced to an individual, the execution
of a new observing project involves tens or more likely hundreds of
people.

While the transition from mainly single-person efforts to team work is
inevitable in the field of computational astrophysics, there are a few
important stumbling blocks that make the transition unnecessarily hard.
One stumbling block is the fact that thesis advisers learned their
trade in a time when simulations were by definition single-person
affairs, and it may be difficult for them to conceive of training
their own students in a radically different way.  

Another important stumbling block is recognition: if a graduate student
would make an important contribution to a team effort of software
writing, it is not at all clear whether that person would get rewarded
sufficiently, especially if the science coming out of the project will
not start up until years after the student receives a PhD.  In
contrast, a student working on, say, the construction of a
gravitational wave detector or a next-generation neutrino detector can
build a promising career well before the first detections occur.

\section{An Example: The Art of Computational Science}

A major unsolved problem, when starting a team effort in setting up a
computational science project, is how to communicate among the project
members.  As mentioned above, this is a problem that has not been solved
yet, either in industry or in academia.  What is clear, however, is
that an {\it Open Source} approach offers the best guarantee for critical
testing and thereby for developing robust code.  With the critical eyes
of anyone in the world interested in the project, major and minor
flaws are likely to be detected soon, a lot sooner than if the project
would be cloaked in secrecy.

We are currently engaged in an attempt to stretch the concept of Open
Source further, to what we call {\it Open Knowledge}.  The main idea
is to provide not only open access to all the computer {\it source
codes} involved in a project, but also to their {\it knowledge base}.
Such a complete disclosure requires documenting the reasoning and
trial and error that went into the production of all relevant codes.
In this way, the background knowledge is made open as well, and others
do not have to repeat the same mistakes -- or, if they wish, they can
critically look at what was labeled as mistakes, to see whether there
may still be some mileage in those attempts, after all.

\begin{figure*}
  \vspace{3cm}
\includegraphics[width=14cm]{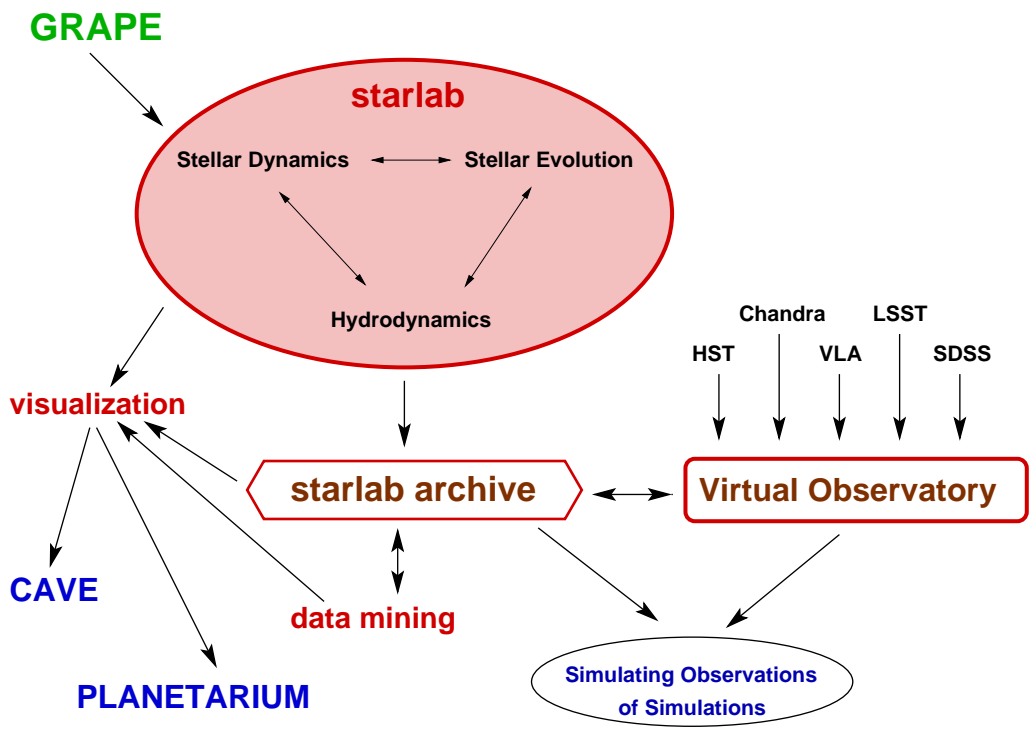}
\caption{An example of a planned framework in which to carry out,
analyze, archive, and present the results of simulations in dense
stellar systems.}
\label{starlab}
\end{figure*}

In other words, besides the {\it what} and {\it how} for any computer
code, we also provide the {\it why}: the motivation for writing it in
the way it was written, within the context in which it was
conceived. This will give the user more appreciation for the
background of the structure chosen, and most importantly, it makes the
codes extensible: it gives the user the ability to easily modify and
extend the codes presented, without running into hidden snags and
unstated assumptions.

And what applies to user friendliness also applies to collaborator
friendliness.  The best way to guarantee that collaborators can work
together coherently on shared pieces of codes is to document the
background knowledge to such an extent that new users and
collaborators alike can pick up that knowledge quickly, without having
to ask their colleagues for any details.

We call our initiative {\it The Art of Computational Science}, ACS for
short, and much more information can be found on our web site (Hut \&
Makino, 2003).  The main novelty in our approach is our use of dialogues,
to provide the `open knowledge' structure.  With our aim of listing far
more background information than is usually done in code comments and
manual pages, we were faced with the challenge to package that extra
information in such a way that it would not just be an unstructured
and boring list of technical details.

After some experimentation, we decided that a dialogue structure would
capture both the spirit and the details of the development process.
We have tested this approach by using some of our dialogue manuscripts
while teaching a couple N-body summer schools, during the last two
years, and we have been very encouraged by the reactions of the
students.  Clearly, our material has filled a gap in the market, in
making explicit what until now has mainly been an oral tradition of
how to set up stellar dynamics experiments from scratch.

\section{Dense Stellar Systems}

Coming back to our main topic, let us list the ingredients that are
needed for the simulation of dense stellar systems.  When a star
cluster is dense enough for individual stars to collide, we need a
hydrodynamics code to model such collisions on a dynamical time
scale.  In addition, we need a stellar evolution code to describe the
subsequent evolution of the merger products, on a thermal time scale
and beyond.  And in order to model the star cluster as a whole, we
need a stellar dynamics code to follow the orbits of all the stars.

In addition to these three ingredients, we need to visualize the
complex processes that occur during the evolution of a dense stellar
system, on all scales of interest, from the system as a whole down to
the details of the modification in, say, nuclear burning inside single
stars and binaries affected by encounters.  We also need a system to
archive long runs, and to make them available for users, theorists as
well as observers.  Finally, to obtain simulation speeds high enough
to model the long-term evolution of a million stars, it is essential
to use the fastest hardware, as has been developed in the GRAPE family
of special-purpose computers.

The three areas of astrophysics mentioned above, stellar dynamics,
stellar evolution and hydrodynamics, are all well developed in their
own right.  Stellar dynamics and stellar evolution each have a history
of half a century of simulations.  The hydrodynamics of stellar
collisions, in comparison, is much less developed.  Each of these
three fields are briefly described below, followed by an equally brief
description of the other three topics, special-purpose computers,
visualization, and an accessible archiving system.  Figure 1, adapted
from the starlab review by Hut (2003), presents a picture of all six
aspects of a framework for a complete dense stellar systems lab.

\section{Stellar Dynamics}

\begin{figure*}
  \vspace{3cm}
\includegraphics[width=13cm]{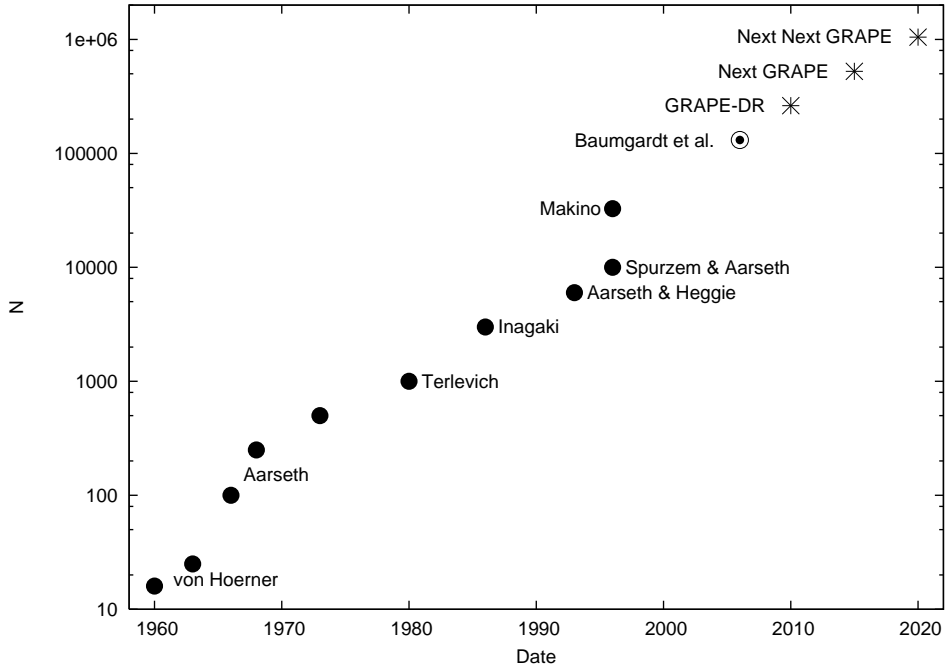}
\caption{Number of particles used in N-body simulations up to and
beyond core collapse, as a function of year of publication.  The dot
labeled Makino refers to calculations using the GRAPE-4.  The
circled dot denotes work in progress, using the GRAPE-6.  The stars
indicate predicted years of publication for runs using computers that
are currently under development or planned to be built.}
\label{n_t}
\end{figure*}

The leading codes to simulate the stellar dynamics of dense stellar
systems are NBODY6 (Aarseth 2003) and Kira (Portegies Zwart {\it et
al.} 2001).  Both codes are freely available.

The published history of computer simulations of the $N$-body problem
starts with von Hoerner (1960), who performed runs with 4, 8, 12, and
16 particles.  Fig.\ \ref{n_t}, adapted from Fig. 3.1 in Heggie \& Hut
(2003), shows achieved and predicted progress over the 60-year time
span since von Hoerner's work.  The GRAPE-DR, which is currently
being built, is expected to be fully operational in 2008, so we expect
to publish results of lengthy runs on the GRAPE-DR by 2010.  The next
two generations of GRAPE computers are expected to finally bring us to
full million-body simulations, all the way through core collapse and
beyond.

Most of the progress in Fig.\ \ref{n_t} comes from Moore's Law, which
for an increase of speed of a factor ten for every five years would
predict an increase in speed of twelve orders of magnitude over the
sixty years depicted here.  Since the cost of N-body calculations in
terms of CPU time scales as $T_{CPU} \propto N^3$ (pair-wise
interactions per crossing time scale as $N^2$, and the relaxation time
grows like $N$), this translates into four orders of magnitude in $N$.
Effectively, the combination of software improvements, mainly by
Aarseth and co-workers, and hardware improvements, mainly by Makino
and co-workers, has resulted in an extra factor of ten in $N$, or a
speed-up of a factor 1,000, beyond the simplest version of Moore's
law.

Note the presence of three distinct roughly linear regimes: 1) von
Hoerner's results, up to 1963; 2) Aarseth's and other's results, up to
1996; 3) the results by Makino and co-workers, starting in 1996.
The jump from 1) to 2) was caused by a software jump in effective
speed, while the the jump from 2) to 3) was caused by the
hardware jump from general-purpose computers to the GRAPE family of
computers.  Note finally that the data points are a bit
heterogeneous: in groups 1) and 3), equal-mass particles were used,
whereas in group 2), mostly unequal mass particles were used, and in
the case of the simulations by Heggie \& Aarseth, some primordial
binaries were included.

Computers available for astronomers around 1960 had speeds measured
in kflops, whereas the result labeled Makino in Fig.\ \ref{n_t} was
obtained with the first computer running at 1 Teraflops, implying an
increase in speed of a factor $10^9$.  The GRAPE-DR is expected to run
at a speed of at least 1 Petaflops.  The next-generation GRAPE, after the
GRAPE-DR, will probably run at a speed of a few tens of Petaflops, at
some date around 2012 or so.  The GRAPE generation after that may
reach a speed of 1 Exaflops, at some time after 2015.

We have estimated in the figure that a speed of 1 Exaflops will enable
us to follow a million-body system of equal-mass particles up to and
beyond core collapse.  Thus the transition from a 16-body to a
million-body system, with an increase in particle number of almost $10^5$,
will have taken a time span of sixty years and a speed increase of a
factor $10^{15}$.

In round numbers, the computing time to core collapse
$t_{sim}$ as a function of $N$ and computer power $P$ can be expressed as:

\begin{eqnarray}
&&t_{sim}(N,P) = \nonumber\\
&&
 \ \ \ \ 0.1 \left({N\over 10^3}\right)^3 \left({1 {\rm Gflops} \over P}\right)
{\rm days} \ \ \ =\nonumber\\
&&
 \ \ \ \ 100 \left({N\over 10^6}\right)^3 \left({1 {\rm Pflops} \over P}\right)
{\rm days}\nonumber
\end{eqnarray}

While N-body simulations are very compute intensive, the memory
requirements $m_{sim}$ are much less so.  A single snapshot for the
masses, positions and velocities for a million-body system can be
stored in less than 100 Mbyte, which implies that we can store $10^4$
such snapshots in a single Tbyte, which will be considered a small
amount of storage by the time we will be able to follow a million-body
system to core collapse.

It is even possible to store the complete history of the types of runs
that are currently performed routinely.  There are about $100N^2$
particle steps needed to reach core collapse in an N-body system,
which implies a storage requirement of the order of $10^4N^2$ bytes or

$$
m_{sim}(N) \  = \ \left({N\over 10^4}\right)^2 {\rm Tbyte}
$$

This estimate can easily be reduced by an order of magnitude or more,
if we store only a small fraction of the individual particle steps,
perhaps in single precision, and use interpolation to estimate the
intermediate positions and velocities.  However, in the presence of
primordial binaries, the above number may increase by one or two
orders of magnitude, depending on how we store the information for the
perturbed binary motion.  The bottom line is that the estimate given
here is a good estimate for the maximum amount of storage needed for
stellar dynamics simulations of dense stellar systems.

\section{Stellar Evolution}

Published results of stellar evolution calculations often take the
form of tracks in the Hertzsprung-Russel diagram, with additional
physical data presented in tabular form.  Such data are very useful
for population synthesis studies.  The simplest models are constructed
from a weighted sum of individual stellar evolution tracks, while more
detailed models incorporate some additional information about binary
stellar evolution.

For dense stellar systems, however, a typical star has a significant
chance to interact and possibly collide with another star during its
lifetime.  In such an environment stars of different ages can exchange
mass, disrupt each other or merge, and their merger products can get
involved in similar interactions; binary stars can encounter single
stars as well as other binaries, where one or more of the stars may
already be a merger product; and so on.  There is no way that one can
anticipate and tabulate all possible multiple-star interactions in
dense stellar systems.  Detailed attempts at population synthesis for
such systems by necessity have to be dynamical, taking into account
the particular ways that stars encounter one another in a given
simulation.

During the last decade, several dynamical population synthesis studies
have appeared ({\it cf.} Portegies Zwart {\it et al.} 2001, Hurley
{\it et al.} 2001).  In these studies, the dynamics of a dense stellar
system is modeled through direct $N$-body integration, while the
stellar evolution is modeled through fitting formulae that have been
obtained from large numbers of individual stellar evolution tracks.
Binary stellar evolution is modeled through the use of semi-analytic
and heuristic recipes (Hurley {\it et al.} 2002).

The studies mentioned above have all used input data obtained with the
stellar evolution code by Eggleton (1973).  As an introduction to this
code, a limited but more structured version has been made available as
the EZ code, by Paxton (2004).  For an introduction to binary stellar
evolution, see Eggleton (2006).  For more background about approaches
to model stellar evolution in the context of dense stellar systems,
see Hut {\it et al.} (2002) and Sills {\it et al.} (2003).

\section{Stellar Hydrodynamics}

Stellar dynamics is perfectly adequate in modeling the motions of
stars as point masses moving under the influence of gravity, even in
dense stellar systems, unless individual stars approach each other to
within a few stellar radii.  When that happens, the internal structure
of the stars has to be taken into account, and we have to switch to a
hydrodynamics module to follow the encounter, which may lead to mass
transfer and even to the merging of two or more stars.  After the dust
has settled, we then have to update the stellar evolution models for
the stars involved, and in case of mergers we will have to construct
new models from scratch, often with highly unusual chemical compositions.

Pioneering investigations of these complex processes have been made by
Sills {\it et al.}  (1997, 2001), where individual collisions and their
aftermath were followed in detail.  The main stages in this process
have recently been automated for the first time, as was reported
during the MODEST-6a workshop in
Lund\footnote{\vbox{http://www.astro.lu.se/\~\,melvyn/} modest6a.html}, Sweden,
in December 2005: Ross Church automatized the collisions, and Evert
Glebbeek automatized the construction of merger models that can be fed
to a stellar evolution program, using the Make Me a Star routine by
Lombardi {\it et al.} (2002).

\section{The GRAPE Family of Special-Purpose Hardware}

Simulations of dense stellar systems involving tens of thousands of
particles would not yet be possible, were it not for the presence of
GRAPE special purpose hardware.  Developed at Tokyo University,
starting in 1989 (Sugimoto {\it et al.} 1990), the GRAPEs made it
possible to simulate core collapse for 32,000 particles already in
mid nineties (Makino 1996), and currently simulations using more than
100,000 particles pas core collapse are routine ({\it cf.} Baumgardt
et al. 2003).

While the earlier GRAPE chips were hardwired to compute gravitational
particle-particle forces, the new GRAPE-DR chip, currently under
development, is moving more in the direction of general-purpose hardware.
It has a SIMD (single-instruction-multiple-data) architecture, which
allows good performance for a large class of scientific applications
(Makino 2006).

\section{Visualization}

In a normal laboratory, it is often difficult or even impossible to
measure some of the physical parameters of interest.  Some areas of
interest may not be large enough to house a detector, and for some
quantities there simply may not exist a detector that is sensitive
enough to measure them.  Some detectors may be too expensive, and so
on.  In contrast, in the virtual laboratory in which simulations take
place, in principle any value of any physical variable can be
determined at any place and time.

This incredible luxury comes at the price, however, namely the price
of writing the proper software to probe the simulations.  Yes, in
principle anything can be determined, but in practice we face the
challenge of writing software tools that are general and flexible
enough to give the user the desirable powers of investigation.

In practice, this price has turned out to be uncomfortably high.
Both in stellar dynamics and in stellar evolution, after decades of
refinement of the basic codes, there still is a surprising lack of
visualization tools needed to probe in detail the way that the basic
codes operate.

Sure, we have packages that display the motions of the stars in
stellar dynamics, and allow the user to zoom in and out at arbitrary
places in space and time.  But wouldn't it be nice to have intelligent
tools that automatically find the most interesting needles in the hay
stack, in the form of, say, close three-body encounters or whatever it
is that one may be interested?  With such tools, one could take the
whole recorded space-time history of a long run, in order to distill
from all that information a few short movies in particular locations
of interest.  Currently, no such tools exist.  Clearly, the development
of these kind of tools would form a great contribution to the field of
the dynamics of dense stellar systems.

Of course, a general visualization system will not be able to recognize
such interesting needles, since it has no domain-specific knowledge.
What is needed is a visualization framework, as an environment that
allows the user to create modules that contain the specific knowledge
needed to find what is really interesting for a particular application.
If the visualization framework provides the supporting infrastructure,
the user will be able to write relatively short programs indicating
what events are of particular interest.  This is similar to the way
any programming language is used: the main complexity lies in the
compiler, which is provided by an outside source, allowing the user to
write far simpler programs for particular applications ({\it cf.} Bischof
et al. 2006).

Similarly, we have ways to plot the trajectory of stars in a
Hertzsprung-Russel diagram, or in a few other diagrams that plot one
physical parameter against another.  But by and large, for most legacy
codes there is a lack of flexible visualization software that would
allow one to easily zoom into what is really going on in particular
shells of interest at particular stages in the evolution of a star.
Here, too, the biggest challenge is not so much to allow access to any
aspect of the data, but rather to perform a type of automatic data
mining, in which an intelligent tool will provide the user with just
the type of data he or she might be interested in.

Recently, a group of astrophysicists and computer scientists have
started to explore how we might develop the kinds of tools needed, in
a series of meetings hosted by Steve McMillan at Drexel University, in
Philadelphia, PA.  Some of the main contributions have been made by
Levy (2003), from the National Center for Supercomputing Applications
at the University of Illinois at Urbana-Champaign, IL, and by Bischof
(2005), from the Rochester Institute of Technology, in Rochester, NY.

\section{A Virtual Observatory for Simulations}

During the last decade, great strides have been made toward the
construction of a virtual observatory.  The basic idea is to allow the
user easy access to a variety of archives for different telescopes,
Earth-based as well as space-based, operating in different wave length
bands, all the way from radio to gamma rays.  Without the need to
learn individual query languages for each data base, transparent
access becomes much easier, and also can be automatized successfully.
For a short recent review, {\it cf.} Djorgovski (2005).

It would be a very natural extension to include archives of simulations
within the scope of a virtual observatory.  An example of a large-scale
simulation in cosmology, that has been made freely available and is now
heavily used by many other researchers, is the Millennium run
(Springel {\it et al.} 2005).

\section{The Need for A Framework}

So far, we have discussed the challenges involved in setting up the
main pieces of an environment in which to run, analyze and
archive the results of simulations of dense stellar systems.  But even
if we have codes to model the stellar dynamics, stellar evolution, and
hydrodynamics, if we have access to a GRAPE system, if we have proper
visualization tools and proper tools to build and interrogate archives,
we are not done yet.

On the contrary, in some sense the main work is just beginning at this
point.  There is the wholly separate challenge of making all these six
pieces talk together and work together.  We need to develop standard
definitions of interfaces between the various software tools, as local
ways to connect individual tools, and we need a framework in
which we can make all these tools operate together in a useful way.

Ideally, we would be able to mix and match bits and pieces from
different software packages, in a type of umbrella environment.  If
this overall framework is designed well enough, we can use it to test
and compare various modules from different codes in different settings.

So far, very little effort has been put into the conception, let alone
the development, of such a framework.  Although many of us have
written tools that we could share with others, in practice we often
wind up rewriting what others have done, because of a lack of ease in
mixing and matching different tools.  Most software tools have little
or no documentation, and often each tool has its own input/output
format and idiosyncratic way of addressing its options through command
line arguments or configuration files.  A central question is: how to
overcome these obstacles without insisting on straitjackets; in other
words: how to leave everyone free to write in their own way and style,
and yet allow an easy sharing of software modules.

A key ingredient of a overarching framework is that we will not insist
on any particular form or format or language or whatever other aspect
of the innards of each tool.  The idea is that all that will be hidden
in a black box, with only the wrapper interface visible to the typical
user.

The main challenge will be to define an extensible software
architecture for a fully equipped computational lab.  An important
step toward that goal is to try to define interfaces and data formats
for communication between modules, while starting off with a series of
toy models, almost trivial black box examples that do little or
nothing but have at least the right interface to be hooked up with
other such modules.  From there on, we can gradually add more and more
astrophysics to the toy models, to let them "grow up" to become real
astrophysics tools.

Ross Church has kindly set up a special email list to discuss this
tool building process.  You can subscribe by going to the MODEST
web site\footnote{http://www.manybody.org/modest.html}  and click on
"mailing list".  You will then find, in the last sentence, a
"here" to click on to bring you to Ross's stellar-discuss page.

\section{The Need for a Flexible Top-Level Language}

Traditionally, legacy codes in (astro)physics have been monolithic,
written as one single huge program, typically in Fortran.  The only
modularity in such an approach is a division in subroutines.  In many
cases, data is passed from one subroutine to another through the use
of common blocks, minimally structured chunks of data that introduce
the danger of connecting everything with everything.  By thus making
far more data visible than is necessary for individual parts of the
whole program, debugging can easily become a nightmare.  Also, it
becomes more complicated to extend such an existing program without
breaking something unrelated somewhere else in the process.

The bottom line is: monolithic programs don't scale gracefully.  The
only reasonable approach toward the construction of a framework for
simulating dense stellar systems, as a virtual laboratory, is to allow
a much more heterogeneous environment.  In such an environment, a
large collection of diverse tools can be used as black boxes, as
mentioned above.  In particular, there is no need to insist on using
the same computer language for different tools.

At the same time, it is important to have a homogeneous and coherent
high-level framework layout with a single suitable language.  Of
course, different groups can set up different laboratories, and as
long as these are designed in a modular way, the various groups can
exchange tools.  Each group can choose its own top-level language, but
it is essential that there is a clear control structure within one
laboratory.  And the easiest way to enforce a homogeneous control
structure is to pick a particular language.

Good candidates for top-level languages are scripting languages, such
as Perl, Python, or Ruby, as well as languages in the Lisp family,
such as Scheme, or a language like Haskell.  What all these languages
have in common is that they are further removed from the hardware
model of computer CPU than more conventional languages.  In the
hierarchy of machine language, assembly language, and more
conventional languages such as Fortran, C, C++, C\# or Java, these
higher-level interpreted languages form yet another stage.

It is much easier to write complex programs in, say, Python than it is
to write the same program in, say, C.  The analogies are that it is
much easier to write a given program in C than it is to write that
program in assembly language; and that it is in turn easier to do that
than to hand code that program in machine language.

The other side of the coin is that a fine-tuned program in assembly
language may be somewhat faster than the same program written in C;
and the C program will certainly be a lot faster than the Ruby
program.  This means that we should only use the highest-level
language in those places where speed is not the bottleneck.

When we are looking for a homogeneous top-level control structure,
speed is obviously not essential, since almost all the work is done in
the most compute-intensive black boxes, where the top-level language
only services as a conductor, orchestrating the whole dance.

So far, use of the newer higher-level languages has only slowly
entered astrophysics.  The main reason for the reluctance to pick up
these new languages is no doubt unfamiliarity.  This in turn creates
two interlocking obstacles: 1) in the middle of everyday
research/teaching/administration pressures, one is not eager to set
aside the time to learn a new type of language, unless its benefits
are overwhelmingly clear; and 2) without getting familiar with a new
type of language, one is unlikely to ever get a real feel for the sort
of advantages that such a new language may bring.

The central problem here is that a lifelong adherence to one
particular language cannot but form a deep single groove that
determines how to think about writing computer programs.  And
it then becomes almost impossible to imagine any alternative.
Without actual praxis in at least one totally different language,
discussions about the pros and cons of switching languages become
unproductive at best, frustrating at second best, and often worse.

Let me try to illustrate this with an analogy.  If someone has been
using Roman numerals for doing arithmetic for many years, (s)he is
probably reluctant to switch to a use of Arabic numerals overnight.
Why, they seem so cumbersome!  There are more different symbols, some
of them look rather similar and you can just imagine how easy it would
be to make mistakes with them.  Besides, you have to learn new tables of
multiplication, tables that seem more complex than the comfortable rules
that you already know, like V*V = XXV or IV*IX = IIII*VIIII =
(I*VIIII) + (I*VIIII) + (I*VIIII) + (I*VIIII) =
VVVV IIII IIII IIII IIII = XX VIII VIII = XX VV VI = XXXVI.

Sure, zealous propagandists of those newfangled Arabic numerals tell
you that you can use them to do miraculous things, like multiplying
MMMDCCCXLIV and MMCMLXXVIII much faster than before, but can you
really take their word for it?  Besides, who would ever have a need 
to add, let alone to multiply, such huge numbers??

\section{An Example: The Maya Open Lab}

An example of a framework for modeling dense stellar systems is the
Maya open lab (Hut \& Makino, 2003).  While it is still in the early
days of being constructed, it already contains well over a thousand
pages of documentation, and a large number of computer codes.  An
example of some of the novel contributions to the Maya lab are N-body
codes that have not only individual time steps, but in addition allow
for individual integration schemes.  While the path of one particle can
be integrated with, say, a leapfrog scheme, another particle can use a
Hermite scheme, yet another particle a fourth-order Runge-Kutta scheme,
or a traditional Aarseth multi-step scheme, and so on.

The Maya lab is currently the main project in our Art of Computational
Science initiative.  And since ACS is based on our notion of `Open
Knowledge', an extension of the idea of `Open Source' as described in
section 5, we call the Maya framework an `Open Lab' because there, too,
the history of and motivation for the construction of the framework is
documented in a very unusual degree of detail.

At the core of the Maya lab, the orchestration of the orbit integration
of the stars and the handshaking between stellar dynamics, stellar
evolution, and stellar dynamics will be taken care of by the Kali code,
currently under construction.  This code is written completely in Ruby,
at least during the prototyping phase, and we will replace the most
compute-intensive parts, where needed, by equivalent modules in C.

We borrowed this name from the Sanskrit {\it kali}, meaning {\it
dark}, as in the {\it kali yuga}, the dark ages we are currently in
according to Hindu mythology.  The same word also occurs in the name
{\it Kali}, for the Hindu Goddess who is depicted as black.  The term
{\it dark} seemed appropriate for our project of focusing on forms of
tacit knowledge that have not been brought to light, so far, and
perhaps cannot be presented in a bright, logical series of statements.
Instead, we expect our dialogues to carry the many less formal and
less bright shades of meaning, that pervade any craft.

As for the name {\it Maya}, this seemed fitting for two reasons, one
connected with Middle America and one with India.  The Maya culture
was very good at accurate calculations in astronomy.  And the word
{\it maya} in Sanskrit has the following meaning, according to the
Encyclopedia Britannica: ``Maya originally denoted the power of
wizardry with which a god can make human beings believe in what turns
out to be an illusion.''  Indeed, a simulation of the heavens is
something virtual, an illusion of sorts, and a considerable feat of
wizardry.

We hope that the Maya open lab will prove to be an adequate and
user friendly framework that can be easily extended for any type of
modeling of dense stellar systems.  Even so, we also hope that others
will construct different frameworks, as part of a friendly competition
in which we can learn from each other and share each others' tools.
Only by trying different approaches will we find out which approach is
most effective for which type of applications.  And as long as we can
agree on interface issues, it should be possible to combine the use
of more than one framework whenever that is desired.

\section{Validation and Verification}

The main reason to carry out and analyze detailed simulations of dense
stellar systems is to compare them critically to observations.  This
process of testing by comparison can be broken down in two steps,
related to validation and verification, two technical terms in
software development.  These engineering terms have the following
official descriptions, as defined by the American Institute of
Aeronautics and Astronautics (AIAA):

\begin{itemize}
\item{Validation: }
    the process of determining the degree to which a model is an
    accurate representation of the real world from the perspective of
    the intended uses of the model. (AIAA G-077-1998)

\medskip

\item{Verification: }
    the process of determining that a model implementation accurately
    represents the developer's conceptual description of the model and
    the solution to the model. (AIAA G-077-1998)

\end{itemize}

If a given model of, say, common envelope evolution is implemented
incorrectly, verification should catch that.  Whether the results
confirm to what observations show, is a matter of validation.

So the first step in testing should be verification, to check that
your program works as intended.  Only when we are comfortable that
that is actually the case, can we do our main task of validation,
of comparing simulations with observations.

Another way of saying this, is:

\begin{itemize}
\item
  verification compares theory and simulations

\medskip

\item
  validation compares simulations and observations

\end{itemize}

Both are important, and very different.  This is a reflection of the
fact that science, which used to be a question of comparing theory and
experiment/observations, now has three, rather than two components:
theory, simulations, experiment/observations.

In fact, there is yet another important intermediate step, one in
which we compare different simulations, based on different
approximations, in order to see how closely their predictions agree
with each other.  This step was discussed at some length at the IAU
Symposium 208 in Tokyo in 2001, resulting in the specification of a
well defined set of initial cluster and stellar parameters (Heggie
2003).  Given the fact that the necessary codes are rather complex,
requiring years of development, so far few groups have been able to
confront this new challenge.  This stands in contrast to the first
collaborative experiment (Heggie {\it et al.}  1998), which was
confined to stellar dynamics (without stellar evolution), and
attracted "entries" from about 10 groups.

\section{Team Work}

At first, modeling dense stellar systems will require the use of
so-called legacy codes.  However, due to the lack of modularity of
these codes, at some point we will have to rewrite those codes.  Given
that the leading codes in use in stellar evolution and stellar dynamics
have a history of decades, how realistic will it be to attempt to
rewrite them?

Of course, in principle a rewrite should take significantly less time
than the time spent originally to write a legacy code, given that we
should have learned from the process of writing the code in the first
place.  In practice, however, this time saving argument is far from
clear.  For one thing, in most cases very little of the original code
writing has ever been documented, and as a result, much of the trial
and error process may have to be repeated.  For another, to make a
much more robust and general code introduces extra requirements, above
and beyond the task of getting something that sort-of works, most but
not all of the time, the typical goals that have been in operation so
far.

Let us estimate how long it would take an expert to produce a complete
rewrite of a legacy code, in a modular and robust and well-documented
way.  To get something working, even while starting from scratch, will
take only a few years.  But to then get the code to the point that it
will include the more fancy additions that have accreted onto the
legacy code will take a few years more.  Most likely, in the process
the original goal of full modularity will have been compromised, more
than once, leading to the need to backtrack a few times, setting
things up again more of less from scratch.  And then there is the
requirement of making everything robust, so that the code will run
under (almost) all conceivable circumstances, without crashing or
grinding to a halt.

All this is likely to take at least ten years, and probably
significantly longer.  If you then add the need to make the code run
efficiently on massively parallel computer clusters, a few more years
can be thrown in easily.  And, last but not least, our expert is
supposed to carry out some scientific projects with the new code, both
to produce scientific results as well as to see whether the code
really performs adequately in cutting-edge research projects.
Therefore, as a round number, 20 person-years may be a realistic
estimate, and if anything perhaps an underestimate.

Now this is based on a direct scaling up of the work that needs to be
done by a single expert.  The question arises: how to develop this
kind of software with a team, say in a 5-year time span?  A lower
limit for the size of the team would be four people, in order to
provide the 20 person-years needed, but that will be a vast
underestimate.  In practice, we will need $\gg 20/5 = 4$ people, since
we will suffer from at least three inefficiency factors, each of which
are $\simgt 2$:

\begin{itemize}
\item 
  each piece of code needs extensive documentation

\item 
  each code writer has to talk extensively with other code writers

\item 
  each code writer is not as brilliant as the single expert

\end{itemize}

If we take all this on face value, we will need at least $(20/4)2^3=40$
persons to produce a complete replacement of a legacy code in five years.
If we want to do this for stellar dynamics, for stellar evolution, and
for stellar hydrodynamics, we have to triple this estimate.  In addition,
we still have to provide similarly robust and modular code for
visualization, archiving and the task of integrating it all in an
overarching framework.  If we make the rather optimistic estimate that
the latter three together require only as much work as the replacement
of a legacy code, we wind up with the requirement of having 160 people
working together to produce a full-fledged, state-of-the-art, modular,
robust, and splendidly documented body of software that can simulate,
in an integrated way, all the physical processes relevant for dense
stellar systems, to the extent that we understand the underlying physics.

\section{Centers for Modeling Dense Stellar Systems}

The analysis above has driven us to a rather large enterprise.
It is hard to say what will be harder: to find 160 individuals suited
to the task, or to find the money to pay them for 5 years; with
overhead for management and a building, we're looking at a project of
order of a hundred million dollars, or twenty million dollars per year
for five years.

Faced with such a demand, one may wonder whether my estimate was not
wildly overblown.  Could the job not be done with far fewer people?
I hope that is the case, and I'd love to hear any good argument in
that direction.  However, such an argument should address specifically
the detailed points I have listed above, and, frankly, I doubt that
such an argument can be constructed.  If anything, I am afraid that I
may have been too optimistic in my estimates.

The problem is that, so far, simulation packages have been written
largely by a single person, or a small group of people, in the
$10^0-10^1$ range.  This is the sole reason that a request to employ
of order $10^2$ people may come as somewhat of a surprise.  But as
soon as we reflect on the infrastructure of (astro)physical research
in general, we see that some of the largest projects employ more than
$10^3$ people, and that projects with $10^2$ are in fact quite common.

One example in astrophysics of a simulation center that falls in the
$10^2$ people category is the ASC/Alliances Center for Astrophysical
Thermonuclear Flashes\footnote{http://flash.uchicago.edu}, based at
the University of Chicago, with widely spread collaborations with many
other universities and research centers.  This center is developing,
maintaining and freely distributing the FLASH code for modeling
thermonuclear flashes, which is now used by many astrophysicists.

What are the prospects that the study of dense stellar systems could
lead to a similar-sized initiative?  The range of topics, from active
galactic nuclei to star forming regions, including the study of
globular and open clusters and planet formation, certainly touches
upon a large fraction of astrophysics research, directly or indirectly.
What would be a realistic way to get such an initiative underway?

Realistically, we will have to start with a group size somewhat
smaller than $10^2$.  As a round number, imagine that we could get
50 people to collaborate, half an order of magnitude than the number
of 160 listed above.  Such a team would not be able to build a dream
laboratory for dense stellar systems in five years, but they still
should be able to make a reasonable start in that direction.  Also,
the time span of five years may be too optimistic anyway: if we
stretch it out to ten years or longer, a team of 50 people may well
be adequate.

There is no need to have these 50 people working together in one center.
The whole idea of code modularity should guarantee that large chunks
of code can be written independently of other large chunks of code.
In principle, a well thought-out, well-balanced and high degree of
modularity could allow 50 people to write code in 50 different locations.
However, such a fine-grained approach strikes me as unrealistic.  My
guess is that productivity will be far higher if people can work in
clusters, with day-to-day communication in a face-to-face way.

One scenario would be to establish 5 centers for producing the tools
for modeling dense stellar systems, with 10 persons actively involved
in tool building at each of these sites, to get a critical mass of
50 people in total.  Given the current distribution of individuals
working in modeling dense stellar systems, there could be one center
in Japan, two in the U.S., and two in Europe.  Each center would need
funding on the level of at least a million dollars per year, depending
on overhead and costs for housing, management, secretarial help, and
the amount of hardware required.  The minimum total yearly cost of
around five million dollars is indeed equal to the funding level of
the FLASH center, mentioned above, so this may be a reasonable estimate,
especially since the cost will be distributed over several countries.

It will neither be realistic nor desirable to carve up the work that
needs to be done over these five centers in an exclusive way.  Each
center will want to keep a working set of tools for all aspects of the
simulations, together with a minimal amount of expertise concerning
those tools.  And a certain amount of competition between the centers
will actually be beneficial, leading to an increased degree of
robustness: those tools that are found to perform best will either be
taken over by other centers, or at least their most relevant design
principles can be incorporated in the further development of tools
elsewhere.

For all this to work, an atmosphere of openness and sharing of code
will be essential, together with regular communication between the
centers.  Fortunately, the MODEST
community\footnote{http://www.manybody.org/modest.html} already has an
excellent track record of stimulating ongoing dialogues between more
than a hundred researchers in the field of dense stellar systems,
through multiple meetings each year for the last several years.
So far, most of the discussions have revolved around plans for the
future, but once the five or so centers are in full swing, the same
communication channels can be used to orchestrate the interactions
between the centers.

\section{Outlook}

Detailed simulations of dense stellar systems, while currently still
on the drawing boards, will become possible over the next five years,
and can be expected to become routine in another five years.  And the
timing is right: simulations in many other areas of astrophysics are
expanding to the point of beginning to overlap with the study of dense
stellar systems.

For example, cosmological simulations are becoming so accurate that
the limiting factor is no longer just the sheer hardware speed or
equivalently the number of particles that can be used in a simulation.
Rather, new bottlenecks are rapidly appearing in the form of the
details of star formation and the behavior of active galactic nuclei,
two examples of dense stellar systems.

Similarly, simulations in galactic dynamics, such as the study of the
collision and subsequent interaction of two galaxies, show that dense
young star clusters are created in the bridges and tails that are
formed during the galactic encounters.  Without modeling the internal
processes that take place in these young dense clusters, the accuracy
of the galactic encounter simulations is inherently limited.

Yet another area of simulations is the formation of planetary systems,
a relevant task given the recent wealth of observations of extrasolar
planets.  Stars are formed in star forming regions, and the formation
of protoplanetary disks is an intrinsic part of the whole star forming
process.  Therefore, the only way to accurately model the formation of
planetary systems is by taking into account, at least to some degree
of realism, the formation of the whole embedding star forming region,
another example of a dense stellar system.

At the heart of the study of dense stellar systems is the study of
stellar evolution, an area that had its heydays in the nineteen
sixties, and became relatively less fashionable in the seventies and
eighties, with a shift to galactic and extragalactic astrophysics,
and in the nineties and the current zeroes, with a shift to precision
cosmology.  However, for the reasons summarized above, I predict that
stellar evolution, through its central role in the study of dense
stellar systems, will once again take center stage in astrophysics,
starting in the next decade, a role that is likely to last for decades.

My conclusion is that young researchers are well advised to learn a
fair amount of stellar evolution, and especially binary star evolution,
where so much is still unexplored.  And making these fields even the
focus of their research is likely to pay off, in almost any application
they will later venture into.

\section*{Acknowledgments}
I thank Jun Makino for many fun discussions about the themes treated
in this paper, and for his comments on this paper.  I also thank
Hans-Peter Bischof, Coleman Miller, Bill Paxton, Gerald Sussman,
John Tromp and Enrico Vesperini for their comments.

\end{document}